\documentclass[aps,pra,twocolumn,superscriptaddress, amsmath,amssymb,10pt]{revtex4-2}
\usepackage[colorlinks=true,linkcolor=blue,urlcolor=blue,citecolor=blue,pdfusetitle]{hyperref}
\usepackage{graphicx}
\usepackage{color}
\usepackage{orcidlink}

\usepackage{amsmath}
\newcommand{\sgn}{\mathrm{sgn}}
\newcommand{\Tr}{\mathrm{Tr}}

\begin{document}
\title{Analytic two-body density-matrix spectrum of the Fermionic Tonks–Girardeau gas}
\author{Abel Rojo-Franc\`{a}s~\orcidlink{0000-0002-0567-7139}}
\email{abel.rojo@oist.jp}
\affiliation{Quantum Systems Unit, Okinawa Institute of Science and Technology Graduate University, Onna, Okinawa 904-0495, Japan.}
\author{Joan Martorell}
\affiliation{Departament de F{\'i}sica Qu{\`a}ntica i Astrof{\'i}sica, Facultat de F{\'i}sica, Universitat de Barcelona, E-08028 Barcelona, Spain}
\author{Francesc Sabater~\orcidlink{0009-0005-1991-4077}}
\affiliation{Technische Universität München, CIT, Department of Computer Science, Boltzmannstraße 3, 85748 Garching, Germany}
\author{Bruno Juli{\'a}-D{\'i}az~\orcidlink{0000-0002-0145-6734}}
\email{bruno@fqa.ub.edu}
\affiliation{Departament de F{\'i}sica Qu{\`a}ntica i Astrof{\'i}sica, Facultat de F{\'i}sica, Universitat de Barcelona, E-08028 Barcelona, Spain}
\affiliation{Institut de Ci{\`e}ncies del Cosmos, Universitat de Barcelona, ICCUB, Mart{\'i} i Franqu{\`e}s 1, E-08028 Barcelona, Spain}
\begin{abstract}
    The two-body density matrix (TBDM) of a quantum many-body system provides information about particle correlations, e.g., pairing. In this work, we derive the complete spectrum of the TBDM of the fermionic Tonks-Girardeau gas using its exact ground-state wavefunction in second quantization. We obtain analytical expressions for all eigenvalues and eigenvectors for an arbitrary number of particles and identify two different types of eigenstates: collective states, formed by superpositions of paired momentum modes, and single-determinant states. We provide a physical interpretation of this structure, explaining the origin of the different eigenstates and their degeneracies. We further derive an asymptotic expansion for the largest eigenvalue, improving on previous estimates and analytically confirming its linear scaling with the particle number, thereby indicating off-diagonal long-range order. For the high-index collective states, we obtain simpler expressions for the eigenvalues and eigenvectors, showing that the eigenvalues decay as $n^{-4}$ and that the eigenvectors become independent of the number of particles. Finally, we discuss the connection between the collective states and an effective description in terms of composite bosons, whose spectral properties suggest a behavior consistent with Tan's contact for interacting bosonic systems.
\end{abstract}
\maketitle
\section{Introduction}
The study of one-dimensional quantum systems is of particular interest because quantum effects are enhanced in lower dimensions~\cite{giamarchi_quantum_2003}. One useful property of these systems is the Bose-Fermi mapping, which relates gases of interacting identical bosons via s-wave interactions to gases of identical fermions interacting via a p-wave potential~\cite{cheon_fermion-boson_1999,girardeau_effective_2004,girardeau_fermi-bose_2003}. By using this mapping, there are two limiting cases with analytical results: the Tonks-Girardeau gas (TG)~\cite{tonks_complete_1936,girardeau_relationship_1960}, where non-interacting fermions are mapped to an infinitely repulsive Bose gas, and the Fermionic Tonks-Girardeau gas (FTG)~\cite{girardeau_static_2005,girardeau_ground_2010}, where we map a non-interacting boson gas to an infinitely attractive interacting fermion gas. For the numerical calculations with s-wave interaction, there exists a large variety of methods and procedures~\cite{busch_two_1998,volosniev_multicomponent_2014,mistakidis_induced_2020,rojo-francas_direct_2022}. But for p-wave interaction, the numerical solution is considerably more challenging, requiring specialized techniques~\cite{kanjilal_nondivergent_2004,camus-sais_coordinate-space_2026}.
On the other hand, it is now possible to create and manipulate one-dimensional systems with few fermions in ultracold-atom laboratories~\cite{serwane_deterministic_2011,pagano_one-dimensional_2014,liao_spin-imbalance_2010,moritz_confinement_2005,boll_spin-_2016}. In these conditions, it is possible to control the parameters of the system, such as the confining trap~\cite{murmann_two_2015}, the interaction between particles~\cite{zurn_fermionization_2012,haller_confinement-induced_2010}, or the number of particles~\cite{wenz_few_2013,jain_programmable_2026}.

Specifically, the Fermionic Tonks-Girardeau gas attracted the attention of the community a couple of decades ago, when evidence of pairing was found~\cite{minguzzi_pairing_2006}. More recently, universal properties of the FTG have been demonstrated~\cite{koscik_universality_2023}. Moreover, in previous work~\cite{sabater_universal_2024,sabater_bardeen-cooper-schrieffer_2025} we have shown explicitly that the wavefunction of the ground state of the FTG has a remarkable paired structure, reminiscent of conventional models of paired systems, but with specific distinguishing features: while pairing usually describes the behavior of fermions around the Fermi surface~\cite{bardeen_theory_1957}, in the FTG with an even number of fermions, all of them form Cooper pairs. In addition, the analytical simplicity of the FTG wavefunction allowed us to derive compact expressions for many of its features~\cite{sabater_bardeen-cooper-schrieffer_2025}. We have already shown that the eigenvalues and eigenfunctions of the one-body density matrix have fully analytic expressions for any number of fermions~\cite{sabater_universal_2024}. 

The properties of the two-body density matrices of quantum many-body systems are a subject of continued interest~\cite{donsa_nonequilibrium_2023,coleman_structure_1963} because they contain the information of the system needed to evaluate the expectation values of two-body operators, either in the Hamiltonian itself or in other observables~\cite{kraus_pairing_2009,sambataro_pairing_2020,alon_fragmentation_2021,cox_exact_2022,gigena_many-body_2021}. One property of special interest is the off-diagonal long-range order (ODLRO), which, in addition, can be used to detect pairing in the system~\cite{yang_concept_1962}. In the case of the FTG, pairing has previously been detected via ODLRO~\cite{girardeau_bosonization_2006}.

In the present work, we derive analytic expressions for the two-body density matrix of the FTG. Exact analytic two-body spectra are available only for selected solvable models~\cite{sambataro_pairing_2020,gigena_many-body_2021}. Thus, the study of the FTG is of significant interest in physics not only in its own right but also because it provides an example of a different paired system whose many properties can be analyzed analytically.

The formal structure of the wavefunction of the FTG can be related to the coboson ansatz~\cite{chudzicki_entanglement_2010,cuestas_fermionic_2020,jimenez_composite-boson_2023}, where one assumes that the system is created by pairs of particles, i.e., composite bosons. 
We use this idea to describe an effective system of composite bosons, and we relate it with the properties of the TBDM of the FTG. Additionally, one can obtain information about the interaction potential using the large-momentum distribution through Tan's contact~\cite{sekino_comparative_2018}. With the results we have obtained, we suggest that the paired particles in the FTG may exhibit an effective short-range interaction, warranting further investigation.

The article is organized as follows. In Sec.~\ref {sec:analytical}, we present the analytical derivation of the Two-Body Density matrix elements for the Fermionic Tonks-Girardeau gas using its second-quantized wavefunction. 
In Sec.~\ref{sec:results}, we present examples of the eigenstates and eigenfunctions, as well as explain their behavior in the limit of a large number of particles and the limit of small eigenvalues.
In Sec.~\ref{sec:interpretation}, we provide a physical interpretation of the TBDM eigenvalue structure. Finally, in Sec.~\ref{sec:conclusions}, we present the main conclusions of this work.
\section{Analytical derivation}\label{sec:analytical}
The Fermionic Tonks-Girardeau gas corresponds to the limit of infinitely attractive spin-polarized fermions with zero-range odd-wave interactions. In that situation, one can obtain the ground state wavefunction at zero temperature using the Bose-Fermi mapping~\cite{girardeau_relationship_1960,cheon_fermion-boson_1999}, where the wavefunction is obtained by antisymmetrizing the wavefunction of non-interacting bosons,
\begin{equation}
    |\Psi\rangle = \prod_{i=1}^N \phi_0(x_i) \prod_{i<j} \sgn(x_i-x_j),
\end{equation}
where $\phi_0(x)$ is the single particle ground state. It has been demonstrated that the eigenvalues of the $r$-body matrix do not depend on the external trap~\cite{koscik_universality_2023}. In our case, we work directly with the Fermionic Tonks-Girardeau wavefunction in second quantization, where for an even number of particles $N$, the wavefunction is expressed as~\cite{sabater_bardeen-cooper-schrieffer_2025}
\begin{align}\label{eq:wf_2q}
    |\Psi\rangle = \sum_{k_1<...<k_{N_p}} \sqrt{p_N(k_1,\ldots,k_{N_p})}P^\dagger_{k_1}\cdots P^\dagger_{k_{N_p}}|0\rangle,
\end{align}
where $a^\dagger_k$ is the creation operator associated with the $k$ index of the one-body density matrix (OBDM) eigenvalues, and $P_k^\dagger\equiv a_k^\dagger a_{-k}^\dagger$. For an odd number of particles, we only need to add a creation operator $a^\dagger_0$ to Eq.~(\ref{eq:wf_2q}). $N_p$ is the number of pairs in the gas; for an even $N$, $N_p = N/2$, and for odd $N$, $N_p = (N-1)/2$. The value of the weights, $p_N(k,\ldots,k_{N_p})$, is 
\begin{equation}\label{eq:coefs}
    p_N(k_1, \ldots,k_{N_p})=\frac{N!}{\pi^{2N_p}} \prod_{i=1}^{N_p} \frac{1}{\bar{k}_i^2},
\end{equation}
where, for a more compact expression, we have used the notation $\bar{k}_i$ to unify those for even and odd numbers of particles, being
\begin{equation}
    \bar{k}_i = 
    \begin{cases}
        k_i-\frac{1}{2},& \text{even $N$}, \\
        k_i, & \text{odd $N$}.
    \end{cases}
\end{equation}
By definition, the density matrix is 
\begin{align}\label{eq:density_matrix}
    \rho = |\Psi\rangle\langle\Psi|=\sum_{\vec{k},\vec{k'}} \sqrt{p_N(\vec{k})p_N(\vec{k'})} |\vec{k}\rangle\langle\vec{k'}|,
\end{align}
where $\vec{k}\equiv k_1,\ldots,k_{N_p}$ and $|\vec{k}\rangle \equiv P^\dagger_{k_1}\cdots P^\dagger_{k_{N_p}}|0\rangle$. 

We are interested in studying the two-body density matrix (TBDM). To obtain the TBDM, one needs to trace the density matrix except for two indices
\begin{equation}
\rho^{(2)}= \left(\begin{matrix} N \\ 2 \end{matrix} \right)\Tr_{3,\ldots,N}(\rho). 
\end{equation}
Using this definition, the trace of the TBDM corresponds to the number of possible different particle pairs, $\Tr\rho^{(2)}=N(N-1)/2$. Note that in Eq.~(\ref{eq:wf_2q}) each $P_{k_i}$ corresponds to a pairwise creation operator, so there are two different indices involved, $k$ and $-k$. We use the fact that each index sign contributes with the same weight, being 
\begin{equation}
    w_{\pm k_i}\equiv w_{k_i}= \frac{(N!)^{1/2N_p}}{\pi} \frac{1}{\bar{k}_i},
\end{equation}
that obeys $p_N(k_1,\ldots,k_{N_p}) = \prod_{i=1}^{N_p} w_{k_i}w_{-k_i}$. 

In the most general case, the reduced two-body density matrix is expressed in terms of the matrix elements of $|k_i,k_j\rangle\langle k_a,k_b|$. Each matrix element can be computed from the result after the trace as,
\begin{align}\label{eq:TBDM}
    &\rho^{(2)}_{i,j,a,b} =\langle\vec{q}_{i,j}|\vec{k}\rangle\langle\vec{k'}|\vec{q}_{a,b}\rangle \\
    &= \sum_{\vec{q}} \sqrt{p_N(k_i,k_j,\vec{q})p_N(k_a,k_b,\vec{q})}|k_i,k_j\rangle\langle k_a,k_b|, \nonumber
\end{align}
where $\vec{q}$ corresponds to $N-2$ elements, not necessarily paired: $\vec{q} \equiv k_1,\ldots,k_{N-2}$, but now each $k_i$ can have any sign, and we are not enforcing having all the pairs ($k_i$,~$-k_i$), a difference from $\vec{k}$. In addition, the notation $|\vec{q}_{i,j}\rangle$ means a Fock vector created with the elements of $\vec{q}$ that does not include the indices $k_i$ and $k_j$. 

Due to the pairwise properties of the wavefunction, the non-vanishing elements of the TBDM are limited: one can identify two different kinds, as can be seen directly from the results of the bra-ket $\langle\vec{q}_{i,j}|\vec{k}\rangle=|k_i,k_j\rangle$; either one can have an unpaired element, i.e., $k_j\ne -k_i$, or a paired state, i.e., $k_j=-k_i$. 

Let us start by explaining the first situation, the unpaired basis elements. If $\langle\vec{q}_{i,j}|\vec{k}\rangle=|k_i,k_j\rangle$ it means that $\vec{q}$ must contain the elements $-k_i$ and $-k_j$, and also $N_p-2$ paired $k_\alpha$. This implies that $\langle\vec{k}'|\vec{q}_{a,b}\rangle=\langle k_i,k_j|$. So, these elements are eigenstates of the TBDM, whose eigenvalues can be computed as
\begin{align}
    \lambda_{i,j}&=\sum_{\vec{q}} \sqrt{p_N(k_i,k_j,\vec{q})p_N(k_i,k_j,\vec{q})} \nonumber \\
    &= \frac{1}{\bar{k}_i^2\bar{k}_j^2}\frac{N!}{\pi^{2N_p}} \sum_{\begin{aligned}k_1<\ldots<k_{N_p-2} \\ k_\alpha \ne k_i\ne k_j \end{aligned}}\frac{1}{\bar{k}_1^2}\cdots \frac{1}{\bar{k}_{N_p-2}^2} ,
\end{align}
where the sum runs over all the values of $k_\alpha$ except repetitions, and the values $k_i$ and $k_j$, but without taking into account the signs. Note that $\lambda_{i,j}$ does not depend on the sign of $k_i$ and $k_j$, so that there are four eigenstates, $|\pm k_i,\pm k_j\rangle$, with the same eigenvalue. 

For odd-$N$, one of the values of $k_i$ can be $k_0$, which is unpaired. This allows one to have components of the TBDM of the form $|k_0,k_i\rangle\langle k_0,k_i|$, that correspond to diagonal states with eigenstate $|k_0,k_i\rangle$. In these cases, the eigenvalues are
\begin{align}
    \eta_{i}&=\sum_{\vec{q}} \sqrt{p_N(k_0,k_i,\vec{q})p_N(k_0,k_i,\vec{q})} \nonumber \\
    &= \frac{1}{\bar{k}_i^2}\frac{N!}{\pi^{2N_p}} \sum_{\begin{aligned}k_1<\ldots<k_{N_p-1} \\ k \ne k_i\end{aligned}}\frac{1}{\bar{k}_1^2}\cdots \frac{1}{\bar{k}_{N_p-1}^2} ,
\end{align}
where $w_0=1$ by definition. Since the eigenvalues do not depend on the sign of $k_i$, both eigenstates $|k_0,\pm k_i\rangle$ have the same eigenvalue.

On the other hand, when we trace with a completely paired $\vec{q}$, the result is also two paired elements. In this situation, we do not have the same restrictions as before, so there are off-diagonal elements in the matrix Eq.~(\ref{eq:TBDM}). However, these states belong to the paired subspace $|k_i,-k_i\rangle\langle k_j,-k_j|$, with matrix element $\mathcal{M}_{i,j}$. These matrix elements can be computed as
\begin{align}
    \mathcal{M}_{i,j} &= \sum_{\vec{q}} \sqrt{p_N(k_i,-k_i,\vec{q})p_N(k_j,-k_j,\vec{q})} \nonumber \\
    & = \frac{1}{\bar{k}_i\bar{k}_j}\frac{N!}{\pi^{2N_p}} \sum_{\begin{aligned}k_1<\ldots<k_{N_p-1} \\ k_\alpha \ne (k_i, k_j) \end{aligned}}\frac{1}{\bar{k}_1^2}\cdots \frac{1}{\bar{k}_{N_p-1}^2} ,
\end{align}
where, in this case, it is possible to have $k_i=k_j$. To obtain the eigenstates of this subspace, it is necessary to diagonalize this matrix $\mathcal{M}$. Unlike the previous eigenstates analyzed, the eigenstates of the matrix $\mathcal{M}$ are not degenerate and are composed of an infinite superposition of states of the paired subspace.
\subsection{Compact expression of the TBDM eigenvalues}
Now, we turn our attention to the summations that share the same structure in all the obtained expressions. First, we start by defining the summation of $R$ elements as
\begin{align}
    \Sigma_R = \sum_{\bar k_1<\ldots<\bar k_R} \frac{1}{\bar k_1^2}\cdots \frac{1}{\bar k_R^2}. 
\end{align}
Using the Girard-Newton formulae, it is possible to obtain the solution of these summations: one can relate the sum of all the variations of sets of $n$ elements in a set of $l$ variables $x$, e.g.,
\begin{align}
    & e_0(x_1,\ldots,x_l) =1, \nonumber\\
    & e_1(x_1,\ldots,x_l) = x_1 + \ldots + x_l,  \nonumber\\
    & e_2(x_1,\ldots,x_l) = \sum_{1\leq i< j \leq l} x_i x_j, \nonumber \\
    & e_n(x_1,\ldots,x_l) =\sum_{1\leq i_1 < \ldots < i_n \leq l} x_{i_1} \cdots x_{i_n},
\end{align}
with the polynomial summations 
\begin{align}
    Q_n(x_1,\ldots,x_l) =\sum_{i=1}^l x_i^n,
\end{align}
as 
\begin{align}
    e_n = (-1)^n  \sum_{\{ m_i\}} \prod_{i=1}^n \frac{(-Q_i)^{m_i}}{m_i!i^{m_i}},
\end{align}
where the sum runs over all nonnegative integer $m_i$ combinations that fulfill $\sum i m_i =n$. 

We relate our sum $\Sigma_R$ to $e_n$ by identifying $x_i$ as $\bar{k}_i^{-2}$ and including all the (infinite) values except the restricted $k_i$ and $k_j$. In our case, the terms $Q_n(i,j)$ -excluding the elements $k_i$ and $k_j$- correspond to 
\begin{align}
    Q_n&(i,j) = \sum_{
        \bar k =1,\,k\ne(k_i,k_j)}^\infty \frac{1}{\bar k^{2n}}  \\
    =& \begin{cases}
        \zeta(2n,1/2)-\frac{1}{(k_i-1/2)^{2n}}-\frac{1-\delta_{i,j}}{(k_j-1/2)^{2n}}, &\text{even $N$},\\
         \zeta(2n,1) -\frac{1}{k_i^{2n}}-\frac{1-\delta_{i,j}}{k_j^{2n}}, & \text{odd $N$},
    \end{cases}   \nonumber
\end{align}
where $\zeta(s,q)=\sum_{k=0}^\infty \frac{1}{(q+k)^s}$ is the Hurwitz zeta function, a generalization of the Riemann zeta function. 

With those definitions, we can write the expressions obtained in a compact form as
\begin{align}\label{eq:eigen_quadr}
    \lambda_{i,j} &= \frac{1}{\bar k_i^2 \bar k_j^2}\frac{N!}{\pi^{2N_p}}\Sigma_{N_p-2}(i,j), \\  \label{eq:eigen_doub}
    \eta_i &= \frac{1}{\bar{k}_i^2}\frac{N!}{\pi^{2N_p}} \Sigma_{N_p-1}(i,i),\\
    \label{eq:eigen_matrix}
    \mathcal{M}_{i,j} &= \frac{1}{\bar k_i \bar k_j}\frac{N!}{\pi^{2N_p}}\Sigma_{N_p-1}(i,j),
\end{align}
where 
\begin{align}
    \Sigma_R(i,j) = (-1)^R  \sum_{\{ m_i\}} \prod_{n=1}^R \frac{(-Q_n(i,j))^{m_n}}{m_n!n^{m_n}}.
\end{align}
Formally, even though they have a different nature, the eigenstates $\eta_i$ have the same value as the diagonal elements of the matrix $\mathcal{M}$; $\eta_i=\mathcal{M}_{i,i}$. Explicit expressions for $\Sigma_R$ for different $R$ values are given in Appendix~\ref{app:explicit_sums}.

In summary, we have obtained a set of eigenvalues for the TBDM. On one hand, the set of $\lambda_{i,j}$ that for each $(i,j)$ pair has four eigenstates,
\begin{equation}
    |v_{\lambda_{i,j}}\rangle = |\pm k_i,\pm k_j\rangle,
\end{equation}
with the same eigenvalue. Only for an odd number of particles do we have the eigenvalues $\eta_i$. For each $i$, there exist two eigenstates,
\begin{equation}
    |v_{\eta_i}\rangle = |k_0,\pm k_i\rangle,
\end{equation}
with that eigenvalue. We associate both $\lambda$ and $\eta$ with a single-determinant eigenstate; i.e., the eigenstates are created from a single Fock state in our basis. Finally, for the paired subspace, we have derived the matrix elements analytically. We call collective states the eigenstates of that subspace, as they correspond to a superposition of paired states.

Even though the matrix of collective states has analytical elements, it is infinite and not directly diagonalizable. But it is still possible to obtain the eigenstates and their eigenvalues. In fact, in Appendix~\ref{app:col_demo}, we show that the eigenvalues of the collective states for $N\geq 4$ are
\begin{equation}\label{eq:col_eigvals}
    \epsilon(z) = \frac{N!}{\pi^4} z^2 \sum_{m=0}^{N_p-2}  \frac{(m+1)z^m}{\pi^{2m}(N-4-2m)!}
\end{equation}
where $z$ is obtained by solving the secular equation $F(z)=N_p-1$, being the secular function for even $N$ 
\begin{equation}
  F(z)=
  \begin{cases}
    \frac{\pi}{2\sqrt{z}}
    \tanh\left(\frac{\pi}{\sqrt{z}}\right), & z>0,\\
    -\frac{\pi}{2\sqrt{-z}}
    \tan\left(\frac{\pi}{\sqrt{-z}}\right), & z<0,
  \end{cases}
  \label{eq:Feven_m}
\end{equation}
whereas for odd $N$,
\begin{equation}
  F(z)=
  \begin{cases}
    \frac{1}{2} \left[\frac{\pi}{\sqrt{z}}\coth\left(\frac{\pi}{\sqrt{z}}\right)-1\right], & z>0,\\
    \frac{1}{2}\left[\frac{\pi}{\sqrt{-z}}\cot\left(\frac{\pi}{\sqrt{-z}}\right)-1\right], & z<0.
  \end{cases}
  \label{eq:Fodd_m}
\end{equation}
Finally, the collective eigenstates are
\begin{equation}\label{eq:eigenstates}
    |v(z)\rangle = \frac{1}{\sqrt{-F'(z)}} \sum_k \frac{\bar{k}}{1+z \bar{k}^2} |k,-k\rangle.
\end{equation}
All the details of the derivation of the eigenstates and eigenvalues of the collective states are presented in Appendix~\ref{app:col_demo}.
\section{Results}\label{sec:results}
\begin{figure}[t]
    \centering
    \includegraphics[width=1.0\linewidth]{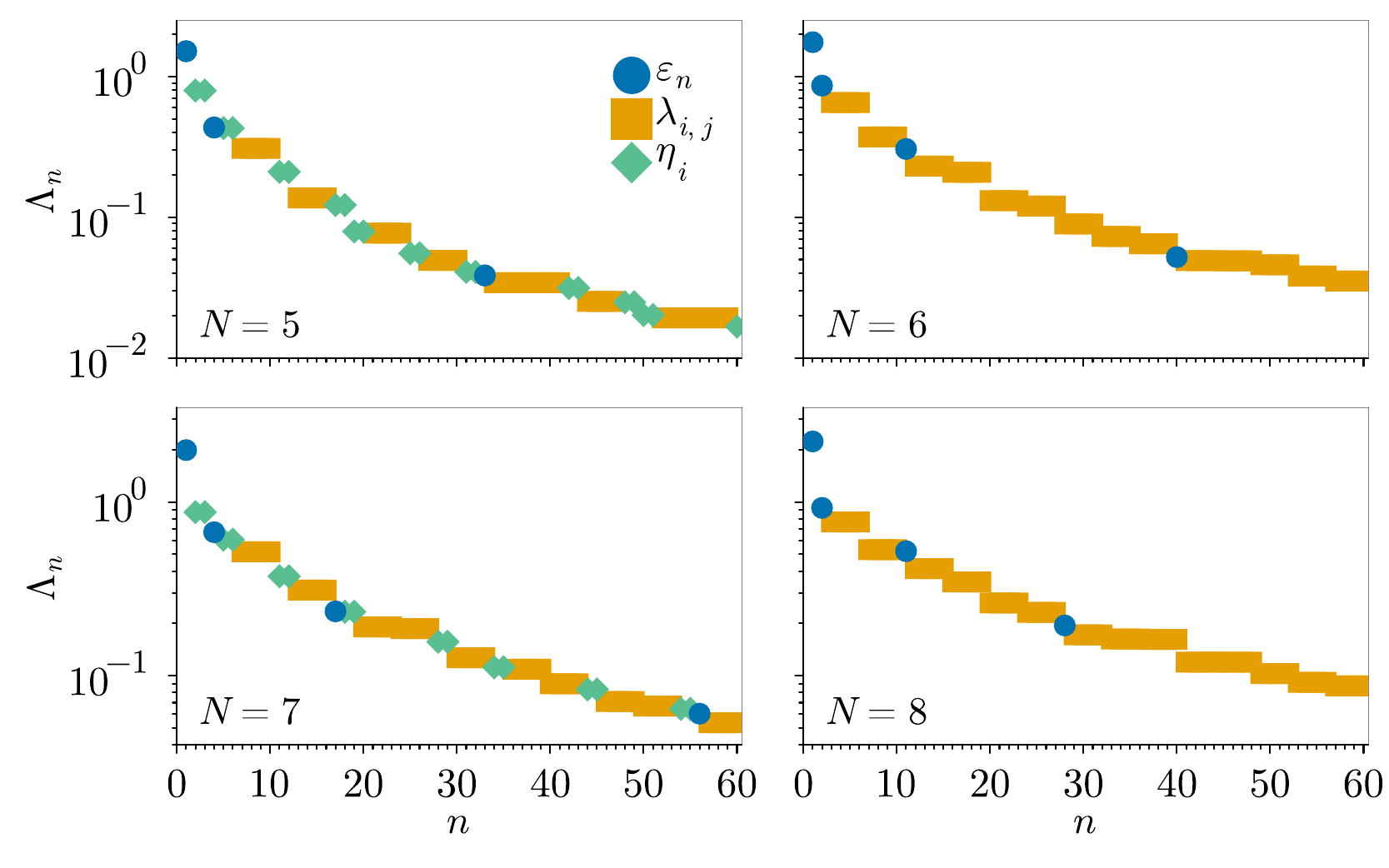}
    \caption{The 60 largest eigenvalues of the TBDM from $N=5$ to $N=8$, showing the collective states $\epsilon_n$ from Eq.~(\ref{eq:col_eigvals}) (no degenerated), and the single-determinant states $\lambda_{ij}$ obtained from Eq.~(\ref{eq:eigen_quadr}) (four times degenerated) and $\eta_i$ obtained from Eq.~(\ref{eq:eigen_doub}) (two times degenerated). The single-determinant state eigenvalues are analytical, and the collective eigenvalues are obtained by computing the root of Eq.~(\ref{eq:Feven_m}) --or Eq.~(\ref{eq:Fodd_m})-- numerically up to numerical precision, and then using the analytical formula. $\Lambda_n$ indicates the total system $n$-th eigenvalue.}
    \label{fig:eigenvalues}
\end{figure}

By solving the roots of $F(z)=N_p-1$ numerically, the calculation of the collective eigenstates and eigenvalues is immediate. With that, we obtained all the eigenstates of the system by analytically computing the single-determinant states, accounting for the corresponding degeneracies. In Fig.~\ref{fig:eigenvalues} we report the larger eigenvalues for different numbers of particles, from $N=5$ to $N=8$. We have cross-checked the results by comparing them with the direct diagonalization of the TBDM in coordinate space, obtaining perfect agreement up to the accuracy of the numerical methods. We have checked the results by diagonalizing the truncated matrix directly using up to $10.000$ basis elements, and by computing the coordinate-space matrix with $250$ points per coordinate, yielding a matrix of size $(250)^2\times(250)^2$. The details of the calculation in coordinate space can be found in Appendix~\ref{app:first_quantization}.

All the analytical derivations are done for $N\geq4$. For completeness, let us present here the cases with fewer particles. In the case of a single particle, there is no TBDM. For two particles, there is only one collective eigenstate with eigenvalue $1$, and the state is $|v(N=2)\rangle=\sum_k \sqrt{2}/[\pi(k-1/2)]|k,-k\rangle$. And in the three-particle case, there is one collective eigenstate with eigenvalue $1$ and eigenvector $|v(N=3)\rangle=\sum_k \sqrt{6}/(\pi k)|k,-k\rangle$, and the collection of the single-determinant states with eigenvalues $\eta_k = 6/(\pi^2k^2)$, associated with the states $|0,\pm k\rangle$.

A general property of the TBDM is that the trace, i.e., the sum of all the eigenvalues, is equal to $N(N-1)/2$, according to the normalization we are using, for any number of particles. We demonstrated that the contribution to the trace of the collective mode states is $\Tr\mathcal{M} = N_p$. Also, the trace contribution of $\lambda_{i,j}$ is $\Tr\rho_{\lambda} = 2N_p(N_p-1)$, and the contribution (only for $N$ odd) of $\eta_i$ is $\Tr\rho_{\eta} = 2N_p$. This result indicates that the total contribution of the single-determinant states becomes dominant as the number of particles increases. The demonstration can be found in Appendix~\ref{app:traces}.

Even though the trace of the single-determinant states dominates, for any $N$, the eigenstate associated with the largest eigenvalue corresponds to a collective mode, i.e., it is not degenerate, and the eigenvector involves an infinite superposition of paired states $|k,-k\rangle$. The behavior of this eigenstate, obtained up to eight particles, is studied in Ref.~\cite{minguzzi_pairing_2006}, where the authors reported a lower bound on the eigenvalue and obtained an approximate behavior of $(N+1)/4$ (note the different normalization in the TBDM), indicating ODLRO. Here, we give an additional interpretation of these eigenstates, associating them with collective states. In addition, with the new formulation, it is possible to obtain the largest eigenvalue with high accuracy for any $N$: by expanding the analytical solution, we obtain the asymptotic value of the largest eigenvalue for a large number of particles, derived in Appendix~\ref{app:asymptote}, as
\begin{equation}\label{eq:eigval_large_N}
    \epsilon_0(N) = \frac{N}{4} +\frac{3}{16} + \frac {19}{64(N-2)}+ \mathcal{O}(N^{-2}).
\end{equation}
This expansion improves the previous approximation from Ref.~\cite{minguzzi_pairing_2006} by a constant term of $-1/16$ and the extra dependence on $N^{-1}$. With the present formulation, we have demonstrated analytically that the largest eigenvalue is linearly dependent on the number of particles, implying that the FTG exhibits ODLRO. 

More recently, in Ref.~\cite{wong_odd-frequency_2026}, they also report an asymptotic value of the largest eigenvalue for this system. Their expression matches our Eq.~(\ref{eq:eigval_large_N}) on the first two terms, but has a discrepancy of $1/64(N-2)$ if one expands their expression in a series in $N$. But, as the expression represents an asymptotic behavior, that discrepancy has almost no impact for large $N$. At the same time, we have provided an exact formula for the eigenvalue, which is extremely efficient for small $N$, making the asymptotic expression unnecessary in these cases.
\subsection{Collective eigenstates}
Now, we turn all of our attention to the collective subspace. 

\begin{figure}[t]
    \includegraphics[width=\linewidth]{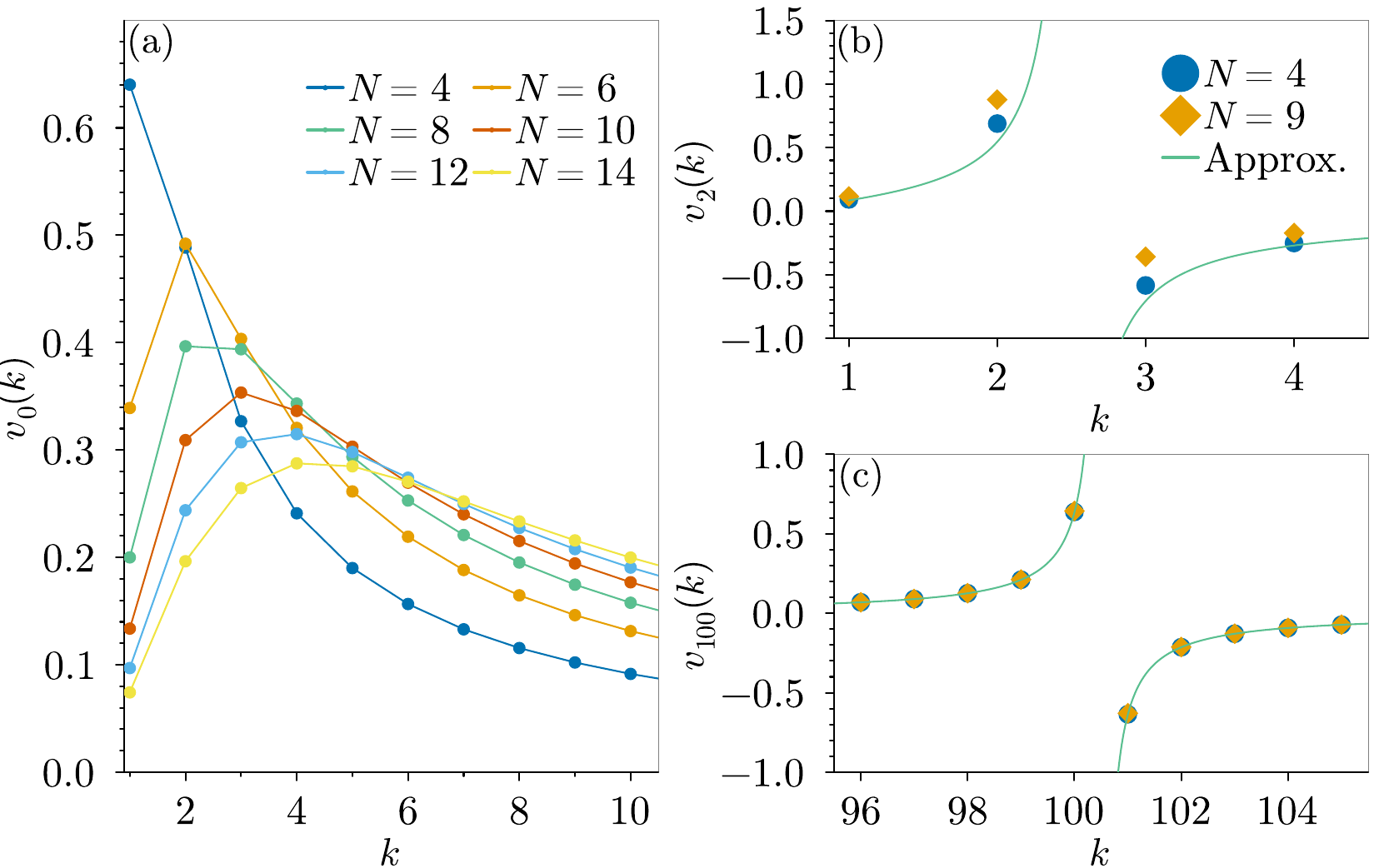}
    \caption{Collective eigenstates of the TBDM. Panel (a) corresponds to the eigenstate with the largest eigenvalue, $v_0$, for different numbers of particles. Panels (b) and (c) correspond to the third (second excited state), $v_2$, and the hundredth excited, $v_{100}$, eigenstates, respectively, for $N=4$ and $N=9$. We also show an approximation for high-index eigenvectors, corresponding to Eq.~(\ref{eq:vec_aprox}), represented by a line for eye guidance. }
    \label{fig:vectors}
\end{figure}
In Fig.~\ref{fig:vectors} we show some collective eigenstates. Specifically, in panel (a), we report the eigenstate associated with the largest eigenvalue ($v_0$) for different numbers of particles. These states contain no nodes and show a clear pattern in the dependence on the number of particles. This eigenstate mixes all the pairs $|k,-k\rangle$ with a probability that decreases for large $k$. For small $k$, there is a maximum contribution to the state with a $k$ value that depends on $N$; as $N$ increases, the maximum is at a larger $k$. In addition, the distribution broadens. It is worth noting that although we reported the eigenstates only for even $N$, the behavior is the same for odd $N$ as well.

In contrast, the remaining eigenstates in the collective sector exhibit a single node. For the $n$-th eigenstate ($v_n$), the node is located between $k=n$ and $k=n+1$ independently of the number of particles $N$. In Fig.~\ref{fig:vectors} (b) and (c), we show the case of $v_2$ and $v_{100}$, respectively. In both cases, we present results for $N=4$ and $N=9$ and observe the same behavior: the node is in the same interval. Moreover, for $v_{100}$, we obtain almost the same result for both $N$, making it impossible to distinguish them in the figure. In fact, as the node $n$ increases, the eigenstates become independent of the number of particles $N$. This can be explained from Eq.~(\ref{eq:eigenstates}), as follows.

First, due to the properties of the secular function, the value of $z$ corresponding to the $n$-th state is between $-1/\bar{n}^2 < z< -1/(\bar{n}+1)^2$, where we used the notation $\bar{n}$ to unify the even and odd $N$. So, for large $n$, one can approximate $z_n\approx -1/(\bar{n}+1/2)^2$. With this value, the normalization term is
\begin{equation}
    \frac{1}{\sqrt{-F'(z_n)}} \simeq \frac{2}{\pi \bar{n}^2}.
\end{equation}
So, the eigenstate can be approximated as 
\begin{equation}\label{eq:vec_aprox}
    |v_n\rangle \simeq \frac{2}{\pi} \sum_k \frac{\bar{k}}{(\bar{n}+1/2)^2-\bar{k}^2}|k,-k\rangle,
\end{equation}
that does not depend on the number of particles. We have compared this approximation in Fig.~\ref{fig:vectors} (b) and (c), obtaining perfect agreement for $v_{100}$, as expected for large $n$.
\subsection{Composite bosons approach}
Obtaining the asymptotic value of the eigenvalues of the collective state subspace of the TBDM is also possible, simply by considering that $N^2|z_n|\simeq N^2/\bar{n}^2 \ll 1$, and so the only term that contributes from Eq.~(\ref{eq:col_eigvals}) is the first term of the sum, obtaining the $n$-th eigenvalue
\begin{equation}\label{eq:sms5}
    \epsilon_n(N) \simeq \frac{N!}{(N-4)!}\frac{1}{\pi^4}\frac{1}{\bar{n}^4}.
\end{equation}
This simple expression reproduces the eigenvalues with high accuracy for large index $n$ and can provide some physics information in the collective space.

First, by considering the wavefunction of the system, Eq.~(\ref{eq:wf_2q}), paired fermions create it, and one can treat each pair $|k,-k\rangle$ as a composite particle, like in the coboson ansatz~\cite{pong_bosonic_2007}. With that in mind, we define the One-Body Density Matrix (OBDM) for these composite particles, where the matrix elements can be obtained in the second quantization as
\begin{equation}
    \rho^{(1)}_{\mathrm{com.}}(i,j) = \langle\Psi|P_{k_i}^\dagger P_{k_j}|\Psi\rangle.
\end{equation}
Writing the explicit fermionic operators, one obtains, in fact, the TBDM in the collective subspace,
\begin{equation}
       \rho^{(2)}_{\mathrm{com.}}(i,j) = \langle \Psi| a_{k_i}^\dagger a_{-k_i}^\dagger a_{-k_j}a_{k_j} |\Psi\rangle \equiv \mathcal{M}_{i,j}.
\end{equation}
The composite particles that describe this system do not correspond to two fermions close to one another; they are Cooper pair-like. This means that the particles are spatially delocalized, but correlated by their momentum. So, to describe composite particles, it is still necessary to use both coordinates: the center of mass and the relative coordinate. 

Nevertheless, if we assume an effective model in which we treat the composite particles as a single particle, certain effects arise. First, by taking into account that the OBDM eigenvalues for an untrapped system correspond to the momentum distribution~\cite{sabater_universal_2024}, the composite particles would have an effective momentum distribution that decays as $n^{-4}$. This decay rate is the same as a bosonic system with contact interaction, following Tan's contact~\cite{sekino_comparative_2018}. This result suggests that the composite particles exhibit an effective interaction. Importantly, as we pointed out before, treating the composite particles as a single object is imprecise in the FTG case. For this reason, the interpretation of the effective interaction can be unphysical, requiring further investigation to identify and analyze it.

\section{Physical interpretation}\label{sec:interpretation}
The second-quantized derivation of the TBDM matrix elements provides a simple way to understand the structure of its spectrum. In particular, it explains why, for even $N$, the spectrum is organized into singlets (non-degenerate collective states) and quadruplets (four-fold degenerate single-determinant states), whereas for odd $N$ an additional family of doublets (two-fold degenerate single-determinant states) appears. The key ingredient is that the FTG ground state is built from pairs of complementary momenta $(k,-k)$ and, for odd $N$, one extra unpaired mode.

A useful picture is to view the TBDM as the two-particle object that remains after tracing out $N-2$ particles. In this picture, the different manifolds are classified by whether the tracing procedure breaks zero, one, or two of the momentum pairs present in the many-body wavefunction.

For even $N$, if the trace breaks two different pairs, say $k_i$ and $k_j$, the remaining two-body state is completely fixed and corresponds to the complementary momenta. This gives rise to the quadruplets, with four possible sign combinations, and explains both their fourfold degeneracy and their non-collective character. Indeed, once the broken pairs are identified, the final state is determined, and no coherent superposition over different pair channels remains.

If, instead, no pair is broken, the remaining two-body state is not reduced to a single product configuration. Rather, it retains the coherent contribution of many $(k,-k)$ channels. This is the origin of the singlets, which are therefore naturally interpreted as collective states.

For odd $N$, the same picture holds, but with one additional possibility. One of the untraced particles can occupy the single-particle ground state $k_0$, while the other belongs to one of the paired modes $k$. In that case, the only nontrivial information is which pair $k$ has been broken. This gives rise to the doublets. As for the quadruplets, these states are not collective, since the tracing procedure singles out a specific pair channel; their two-fold degeneracy follows from the two sign choices $\pm k$.

Therefore, the second-quantized formalism not only reproduces the observed classification of the spectrum, but also provides a clear physical interpretation of it: quadruplets and doublets arise from sectors in which one can identify the broken pair structure explicitly, whereas singlets arise from the sector in which no pair is broken and the coherent superposition over all Cooper-pair channels is preserved. Explicit examples for $N=3$ and $N=4$ are given in Appendix~\ref{app:projection_examples}.
\section{Conclusions}\label{sec:conclusions}
In this article, we analyzed the properties of the Two-Body Density Matrix of a Fermionic Tonks-Girardeau gas by using the exact second-quantization formulation of the wavefunction. We obtained analytical expressions for the eigenvalues and eigenvectors of an arbitrary number of particles, distinguished between collective states and single-determinant eigenstates, and provided a physical interpretation for each kind of state.

We focused on the collective states, as the underlying physics is richer. Firstly, the largest eigenvalue corresponds to a collective state, and its value scales linearly with the number of particles, indicating ODLRO. Even though the existence of ODLRO was described in the FTG, we have now obtained an analytical demonstration of this scaling of the first eigenstate. 

On the other hand, the collective subspace can be understood as the one-body density matrix of the coupled fermions that creates effective composite particles, similar to the coboson ansatz. We have used this interpretation to suggest a possible effective interaction between the composite particles, but requiring further investigation.

\begin{acknowledgments}
We thank Anna Minguzzi for a useful discussion.

This work has been funded by the Okinawa Institute of Science and Technology Graduate University. We are grateful for the Scientific Computing and Data Analysis (SCDA) section of the Research Support Division at OIST.
Additional funding comes from Grant PID2023-147475NB-I00 funded by MICIU/AEI/10.13039/501100011033 and FEDER, UE, by grants 2021SGR01095 from Generalitat de Catalunya, and by Project CEX2019-000918-M of ICCUB (Unidad de Excelencia María de Maeztu).

\end{acknowledgments}
\appendix
\section{Explicit expressions for \texorpdfstring{$\Sigma_R(i,j)$}{SigmaR(i,j)}}\label{app:explicit_sums}
Here, we explicitly show the expressions of $\Sigma_R(i,j)$ for small values of $R$. Note that for several particles $N$, the $\Sigma_R(i,j)$  that are involved are $R=N_p-1$ and $R=N_p-2$. This means that for $N=3$ we will need only $\Sigma_0(i,j)=1$. For $N=4$ and $N=5$, we will need $\Sigma_1(i,j)$ (and $\Sigma_0(i,j)$). For $N=6$ and $N=7$, we need $\Sigma_2(i,j)$ and $\Sigma_1(i,j)$, and so on.

To shorten the notation, we omit the dependence on $k_i$ and $k_j$, using $\Sigma_R(i,j) \equiv \Sigma_R$ and $Q_n(i,j) \equiv Q_n$ instead. The explicit values until $R=6$ are
\begin{align}
    \Sigma_0 =& 1, \displaybreak[1] \\
    \Sigma_1 =& Q_1, \displaybreak[1] \\
    \Sigma_2 =& \frac{1}{2!}\left(Q_1^2-Q_2\right), \displaybreak[1] \\
    \Sigma_3 =& \frac{1}{3!}\left( Q_1^3-3Q_1Q_2+2Q_3\right), \displaybreak[1] \\
    \Sigma_4 =& \frac{1}{4!}\left(Q_1^4- 6Q_1^2Q_2 +3Q_2^2+8Q_1Q_3-6Q_4  \right), \displaybreak[1] \\
    \Sigma_5 =& \frac{1}{5!}\left( Q_1^5 - 10 Q_1^3 Q_2 + 15 Q_1 Q_2^2 + 20 Q_1^2 Q_3 \right. \nonumber& \\
     &\left. - 20 Q_2 Q_3 - 30 Q_1 Q_4 +24 Q_5 \right) , \displaybreak[1] \\
    \Sigma_6 =& \frac{1}{6!}\left( Q_1^6 - 15 Q_1^4 Q_2 + 45 Q_1^2 Q_2^2 - 15 Q_2^3 \right. \nonumber&\\
    &+ 40 Q_1^3 Q_3 - 120 Q_1 Q_2 Q_3 + 40 Q_3^2 \nonumber \\
    &-90 Q_1^2 Q_4 + 90 Q_2 Q_4  \nonumber \\
    &\left.+ 144 Q_1 Q_5 - 120 Q_6 \right).
\end{align}
%
%
\section{Collective eigenstates demonstration}\label{app:col_demo}
The present derivation was originally generated using ChatGPT Sol 5.6 and subsequently analyzed and rewritten by the authors to make it clearer and more natural.

To demonstrate that the eigenstates and eigenvalues of the collective modes are given by Eq.~(\ref{eq:col_eigvals}), let us define a new notation. Now, we define $\Sigma_R(i,...,j)$ as the sum of $R$ elements $1/\bar{k}^2$ except for $k_i,...,k_j$. If there are no indices explicitly, e.g., $\Sigma_R$, it corresponds to a sum of all the $k$ values (without repetitions). For simplicity, we define $a_k\equiv 1/\bar{k}^2$. With that, the matrix element can be written as
\begin{equation}
    \mathcal{M}_{i,j} = \frac{N!}{\pi^{2N_p}}\sqrt{a_ia_j}\Sigma_{N_p-1}(i,j).
\end{equation}
Now, we define the secular function
\begin{equation}
    F(z) = \sum_{k=1}^\infty \frac{a_k}{a_k+z},
\end{equation}
and we consider the $k$-component of the eigenvector ansatz
\begin{equation}
    v_k(z) = \frac{\sqrt{a_k}}{a_k+z}.
\end{equation}
We also define for convenience $y_k(z) = a_k/(a_k+z)$.

With these ingredients, we compute the action of the matrix on the vector,
\begin{equation}
    (\mathcal{M}v)_i  = \frac{N!}{\pi^{2N_p}}\sqrt{a_i} \left( y_i\Sigma_{N_p-1}(i) + \sum_{j\ne i} y_j \Sigma_{N_p-1}(i,j) \right),
\end{equation}
separating the diagonal terms explicitly. We define $S_i \equiv (\mathcal{M}v)_i / (\sqrt{a_i}N!/\pi^{2N_p})$. We also define $r\equiv N_p -1$, obtaining a more compact expression,
\begin{equation}
    S_i = y_i \Sigma_{r}(i) + \sum_{j\ne i} y_j \Sigma_r(i,j).
\end{equation}
Take into account the definition of $\Sigma_r$, being a sum of $r$ element product,
\begin{equation}
    \Sigma_r = \sum_{k_1<...<k_r} a_{k_1} ... a_{k_r},
\end{equation}
where it can include the condition of excluding some index $k_i$. We can rearrange the sum order as,
\begin{equation}\label{eq:si1}
    S_i = \sum_{k_1<...<k_r | k_n\ne i} a_{k_1}...a_{k_r}\left( y_i + \sum_{j \ne \{k_n,i\} } y_j \right),
\end{equation}
where the condition in the first sum indicates a summation over all the values of $k$ without repetitions, and none of the $k_n$ ($n=1,...,r$) can take the value $i$. The second sum indicates that the index $j$ differs from all values of $k_n$ and also from $i$. 

Now we impose the condition $F(z) = r$, meaning that $\sum_j y_j(z) = r$. With that condition, we can obtain the value of the internal sum as
\begin{align}
    y_i+ \sum_{j\ne \{k_n,i\}} y_j &= \sum_j y_j - \sum_{j=\{k_1,..,k_r\}}y_j  \nonumber \\
    &=r - \sum_{j=\{k_1,..,k_r\}}y_j = \sum_{j=\{k_1,..,k_r\}}(1-y_j) \nonumber \\&=z \sum_{j=\{k_1,..,k_r\}}\frac{1}{a_j+z}.
\end{align}
We use this result in Eq.~(\ref{eq:si1}),
\begin{align}\label{eq:Si2}
    S_i &=  \sum_{k_1<..<k_r | k_n\ne i} a_{k_1}... a_{k_r} z \sum_{j=k_n} \frac{1}{a_j+z} \nonumber \\
    &= z\sum_{j\ne i}\frac{1}{a_j + z} a_j \sum_{k_1<...<k_{r-1}|k_n\ne i\ne j} a_{k_1}...a_{k_{r-1}} \nonumber \\
    &= z\sum_{j\ne i} y_j \Sigma_{r-1}(i,j).
\end{align}
Even though this last expression is compact, we start expanding and recovering a similar expression as in the first line of Eq.~(\ref{eq:Si2}),
\begin{equation}
    S_i = z \sum_{k_1<...<k_{r-1}|k_n\ne i} a_{k_1}...a_{r-1} \sum_{j\ne k_n \ne i} y_j,
\end{equation}
and now, the last sum can be obtained as
\begin{align}
    \sum_{j\ne k_n \ne i} y_j &=  \sum_j y_j - \sum_{j = k_n} y_j -y_i \nonumber \\ &= r -\sum_{j = k_n} y_j -y_i = 1+ \sum_{j = k_n} (1-y_j) -y_i,
\end{align}
as there are $r-1$ values of $k_n$ . With that, we obtain 
\begin{align}\label{eq:si2}
    S_i  =& z \sum_{k_1<...<k_{r-1}|k_n\ne i} a_{k_1}...a_{r-1} \left((1-y_i)+ \sum_{j = k_n} (1-y_j)\right) \nonumber \\
    =& z(1-y_i)\Sigma_{r-1}(i)  \nonumber \\
    & +z^2\sum_{k_1<...<k_{r-1}|k_n\ne i} a_{k_1}...a_{r-1} \sum_{j=k_n} \frac{1}{a_j+z} \nonumber \\
    =& z(1-y_i)\Sigma_{r-1}(i) +z^2\sum_{j\ne i} y_j \Sigma_{r-2}(i,j).
\end{align}
By doing a similar procedure on the last term until obtaining $\Sigma_0=1$, one can obtain the compact expression,
\begin{equation}
    S_i = \sum_{m=1}^r (m-y_i) \Sigma_{r-m}(i)z^m,
\end{equation}
where the factor $m$ in the $m-y_i$ is because in each iteration of Eq.~(\ref{eq:si2}), there is one less element in the sum, and so, an extra factor appears out of the sum.

Now, we compute 
\begin{equation}\label{eq:si3}
    (a_i+z)S_i = \sum_{m=1}^r (m-y_i)(a_i+z)\Sigma_{r-m}(i)z^m,
\end{equation}
where we use 
\begin{equation}
    (m-y_i)(a_i+z) = m(a_i+z) -a_i= a_i(m-1) +mz.
\end{equation}
Inserting that result into Eq.~(\ref{eq:si3}),
\begin{align}
    (a_i+z)S_i &= \sum_{m=1}^r (m-1)a_i\Sigma_{r-m}(i)z^m + \sum_{m=1}^r mz\Sigma_{r-m}(i)z^m \nonumber \\
    &=\sum_{m=1}^{r-1} ma_i\Sigma_{r-m-1}(i)z^{m+1} + \sum_{m=1}^r m\Sigma_{r-m}(i)z^{m+1} \nonumber \\
    &= \sum_{m=1} ^ r mz^{m+1}(a_i \Sigma_{r-m-1}(i) + \Sigma_{r-m}(i)),
\end{align}
where we have used that $\Sigma_{-1} \equiv 0$. Then, by using the property $\Sigma_q = \Sigma_q(i) + a_i\Sigma_{q-1}(i)$, we obtain the expression
\begin{equation}\label{eq:si4}
    (a_i+z)S_i = z^2\sum_{m=1}^r m\Sigma_{r-m} z^{m-1} = z^2 R_r'(z),
\end{equation}
where we have defined 
\begin{equation}
    R_r(z) = \sum_{m=0}^r \Sigma_{r-m}z^m.
\end{equation}
By replacing the definition of $S_i$ in Eq.~(\ref{eq:si4}), we obtain
\begin{align}
    &(\mathcal{M}v(z))_i = z^2R_r'(z) \frac{N!}{\pi^{2N_p}}\frac{\sqrt{a_i}}{a_i+z}\\
    &\sum_j \mathcal{M}_{ij} v_j(z) = \epsilon(z) v_i(z),
\end{align}
where 
\begin{equation}
    \epsilon(z) = \frac{N!}{\pi^{2N_p}} z^2 R'_{N_p-1}(z),
\end{equation}
demonstrating that the vector $v(z)$ is an eigenstate with eigenvalue $\epsilon(z)$ when $F(z) = N_p-1$.

The normalized vector can be obtained as
\begin{equation}
    -F'(z) = \sum_k \frac{a_k}{(a_k+z)^2}=\sum_k |v_k(z)|^2,
\end{equation}
and so, the normalized eigenvectors are
\begin{equation}
    |v(z)\rangle = \frac{1}{\sqrt{-F'(z)}} \sum_k \frac{\bar{k}}{1+z \bar{k}^2} |k,-k\rangle.
\end{equation}
By definition, the secular function has a closed form. For even $N$ corresponds to Eq.~(\ref{eq:Feven_m}), whereas for odd $N$ corresponds to Eq.~(\ref{eq:Fodd_m}). Finally, it is possible to obtain a closed form for $\Sigma_r$. First, notice that
\begin{equation}
    \prod_j^\infty (1+a_jt) = \sum_{q=0}^\infty \Sigma_q t^q.
\end{equation}
At the same time, using the Euler functions for the hyperbolic cosine and sine, we obtain 
\begin{align}
    &\prod_{k=1}^\infty \left(1+\frac{t}{(k-1/2)^2} \right) = \cosh (\pi\sqrt{t}),  &  \text{even $N$},\\ 
    &\prod_{k=1}^\infty \left(1+\frac{t}{k^2} \right)= \frac{\sinh(\pi\sqrt{t})}{\pi\sqrt{t}}, &  \text{odd $N$}.
\end{align}
By expanding the hyperbolic functions and identifying term by term, it is immediate to obtain that
\begin{equation}\label{eq:Sigma_n_sum}
    \Sigma_n = 
    \begin{cases}
        \frac{\pi^{2n}}{(2n)!}, & \text{$N$ even}, \\
        \frac{\pi^{2n}}{(2n+1)!}, & \text{$N$ odd}.
    \end{cases}
\end{equation}
With that, we can obtain the final result for the eigenvalues of Eq.~(\ref{eq:col_eigvals}).
%
%
\section{Calculation of the TBDM in first quantization formalism}\label{app:first_quantization}
In this Appendix, we derive a first-quantization formulation for computing the TBDM. Results obtained with this method have been used to cross-check the completeness of the analytical results.

We define the two-body density matrix as
\begin{align}\label{eq:tb0}
    \rho^{(2)}&(x_1,x_2; x'_1,x'_2) \equiv \frac{N ( N-1)}{2} \int \ \Psi^*(x_1,x_2,x_3,...,x_N)\nonumber \\
    &\times \Psi(x'_1,x'_2,x_3,...,x_N) \ dx_3 ... dx_N ,
\end{align}
where $\Psi(x_1,..x_N)$ is the wavefunction.

The eigenvalues, $K_{\kappa}$, and eigenfunctions, $u_{\kappa}$,  of the TBDM are the solutions of:
\begin{align}\label{eq:tb01}
    K_{\kappa}\ u_{\kappa}(x_1,x_2) =  \int \ dx'_1 dx'_2\ \rho^{(2)} (x_1,x_2; x'_1,x'_2)\ u_{\kappa}( x'_1,x'_2)  ,
\end{align}
where $\kappa$ indicates the set of indices needed to characterize the solutions. In general it is known that $\Tr\rho^{(2)} = \sum_{\kappa} K_{\kappa} = N(N-1)/2$.

From Eqs.~(4) to (6) of Ref.~\cite{koscik_universality_2023}, particularizing to $p=2$, we find that 
\begin{equation}\label{eq:dm1}
    \rho^{(2)}_{KS}(x_1,x_2; x'_1,x'_2) \equiv \frac{{\cal P}^{N-2}}{L^2} \sgn(x_1-x_2) \sgn(x'_1-x'_2),
\end{equation}
where
\begin{equation}
    {\cal P} \equiv \int dz\, \sgn (x_1-z) \sgn(x'_1-z) \sgn(x_2-z) \sgn(x'_2-z) ,
\end{equation}
and the eigenvalue equation is
\begin{equation}\label{eq:dm01}
      \int \ dx'_1 dx'_2 \ \rho^{(2)}(x_1,x_2;x'_1,x'_2) u_{\kappa}(x'_1,x'_2) = \lambda^{(2)}_{\kappa} u_{\kappa}(x_1,x_2) .  
\end{equation}
In addition we have introduced a factor $N(N-1)/2$ in the definition of $\rho^{(2)}$, Eq.~(\ref{eq:tb0}). 

To get a more manageable expression for ${\cal P}$ in Eq.~(\ref{eq:dm1}) above, we proceed in several steps:
\begin{align}
    \int_0^1 &\ dz\, \sgn(x-z) = \int_0^x \ dz - \int_x^1 \ dz  = 2x-1 \\
      \int_0^1 & \ dz\, \sgn(x_1-z)\, \sgn(x_2-z) = \label{eq:dm2}\\
      &\int_0^{x_1} \ dz - \int_{x_1}^{x_2} \ dz  + \int_{x_2}^1 \ dz  = 2x_1 -2x_2 + 1 \nonumber ,
\end{align}
where we have assumed $x_1 < x_2$ . Obviously, if instead we had $x_1 > x_2$, then the final result in Eq.~(\ref{eq:dm2}) would have $x_1$ and $x_2$ interchanged. To include both cases in a single expression:
\begin{equation}\label{eq:dm3}
    \int_0^1  dz \,\sgn(x_1-z)\, \sgn(x_2-z) = 1 -2 |x_2-x_1| .
\end{equation}
The next step is to consider the integral involving three signs: Assuming that
$x_1 < x_2 < x_3$ one finds
\begin{align}\label{eq:dm4}
&& \int_0^1 \, dz \, \sgn(x_1-z)\, \sgn (x_2-z)\, \sgn(x_3-z) \nonumber \\
&&= ... = 2x_1-2x_2+2x_3 -1 .
\end{align}
And finally, assuming that $x_1 <x_2 <x_3 < x_4$:
\begin{align}\label{eq:dm5}
&& \int_0^1 \, dz \, \sgn(x_1-z)\, \sgn (x_2-z)\, \sgn(x_3-z)\, \sgn(x_4-z) \nonumber \\
&&= ... = 2x_1-2x_2+2x_3 -2x_4 +1 .
\end{align}
To generalize this result to other orderings, we do this in two steps
: 1) given a set $x_1, x_2, x_3, x_4$  , reorder them in increasing order: $x_1^R < x_2^R < x_3^R < x_4^R$, and 2) the desired result is then
\begin{equation}\label{eq:dm6}
    {\cal P} = 2(x_1^R-x_2^R+x_3^R-x_4^R) + 1  .
\end{equation}
And to use the notations in Eq.~(\ref{eq:dm1}) replace $x_3$ and $x_4$ by $x'_1$ and $x'_2$. 

In this way, it is straightforward to write code to compute the TBDM. 
%
%
\section{Trace calculations}\label{app:traces}
The trace values of the reduced subspaces are straightforward calculations using the previous results. We need the results of the single-determinant eigenvalues, Eq.~(\ref{eq:eigen_quadr}) and (\ref{eq:eigen_doub}), the collective matrix Eq.~(\ref{eq:eigen_matrix}), and the result obtained in Eq.~(\ref{eq:Sigma_n_sum}). Computing the trace of the reduced density matrix is equivalent to summing all the diagonal elements in that subspace. 

We start by computing the trace of the $\eta$ and $\mathcal{M}$, as they are equal ($\mathcal{M}_{i,i} = \eta_i$). So, the trace is
\begin{align}
    \Tr \mathcal{M} =& \frac{N!}{\pi^{2N_p}}\sum_{i} \frac{1}{\bar k_i^2}\sum_{k1<\dots<k_{N_p-1}}\frac{1}{\bar k_1^2}\cdots\frac{1}{\bar k_{N_p-1}^2} \nonumber \\
    =&\frac{N!}{\pi^{2N_p}}N_p\sum_{k1<\dots<k_{N_p}}\frac{1}{\bar k_1^2}\cdots\frac{1}{\bar k_{N_p}^2} \nonumber \\
    =& N_p \frac{N!}{\pi^{2N_p}}\Sigma_{N_p} \nonumber \\
    =&N_p,
\end{align}
where the factor $N_p$ in the first step is because it corresponds to the number of possible permutations of $\bar k_i$ in the $N_p$ elements. Computing $\Tr\rho_\eta$ only requires multiplying it by $2$, as each eigenvalue has a twofold degeneracy, so
\begin{equation}
        \Tr\rho_\eta = 2N_p
\end{equation}

The case of the remaining subspace can be obtained similarly, 
\begin{align}
    \Tr \rho_\lambda =& 4\frac{N!}{\pi^{2N_p}} \sum_{i<j} \frac{1}{\bar k_i^2}\frac{1}{\bar k_j^2}\sum_{k_1<\dots<k_{N_p-2}} \frac{1}{\bar k_1^2}\cdots \frac{1}{\bar k_{N_p-2}^2} \nonumber \\
    =& 2\frac{N!}{\pi^{2N_p}} \sum_{i,j} \frac{1}{\bar k_i^2}\frac{1}{\bar k_j^2}\sum_{k_1<\dots<k_{N_p-2}} \frac{1}{\bar k_1^2}\cdots \frac{1}{\bar k_{N_p-2}^2} \nonumber \\
    =& 2\frac{N!}{\pi^{2N_p}} (N_p-1) \sum_{i} \frac{1}{\bar k_i^2}\sum_{k_1<\dots<k_{N_p-1}} \frac{1}{\bar k_1^2}\cdots \frac{1}{\bar k_{N_p-1}^2} \nonumber \\
    =& 2\frac{N!}{\pi^{2N_p}} (N_p-1)N_p\sum_{k_1<\dots<k_{N_p}} \frac{1}{\bar k_1^2}\cdots \frac{1}{\bar k_{N_p}^2} \nonumber \\
    =& 2\frac{N!}{\pi^{2N_p}} (N_p-1)N_p \Sigma_{N_p} \nonumber \\
    =& 2N_p(N_p-1).
\end{align}
%

%
%
\section{Asymptotic value of the largest eigenvalue}\label{app:asymptote}
The present derivation was originally generated using ChatGPT Sol 5.6 and subsequently analyzed and rewritten by the authors to make it clearer and more natural.

The largest eigenvalue corresponds to the solution for the positive $z$ of Eq.~(\ref{eq:Feven_m}) and (\ref{eq:Fodd_m}). To simplify the notation, we introduce $x_0\equiv \pi/\sqrt{z}$ and $A\equiv N-2$. With that, we obtain the secular equations
\begin{equation}
\begin{cases}
    x_0 \,\mathrm{tanh}(x_0) = A & \text{$N$ even},\\
    x_0\, \mathrm{coth}(x_0)=A & \text{$N$ odd}.
    \end{cases}
\end{equation}
Using the exponential form of the hyperbolic functions and using the transformation $y=2(x_0-A)$, we obtain
\begin{equation}
    \begin{cases}
        y\mathrm{e}^y - y\mathrm{e}^{-2A}=4A\mathrm{e}^{-2A} & \text{$N$ even},\\
        y\mathrm{e}^y + y\mathrm{e}^{-2A}=-4A\mathrm{e}^{-2A} & \text{$N$ odd}.
    \end{cases}
\end{equation}
The solution of this equation corresponds to the $r$-Lambert generalization \cite{mezo_generalization_2017}, defined as
\begin{equation}
    x \mathrm{e}^x +r x =a,
\end{equation}
with solution
\begin{equation}
    x= W_r(a).
\end{equation}
In our case, $r=(-1)^{A+1}\mathrm{e}^{-2A}$ and $a=(-1)^{A}4A\mathrm{e}^{-2A}$, and the Taylor series of the function $W_r(a)$ is
\begin{equation}
    W_r(a)=4A\sum_{n=1}^\infty (-1)^{An}\frac{L_{n-1}^1(4An)}{n} \mathrm{e}^{-2An},
\end{equation}
where $L_n^a(x)$ is the generalized Laguerre polynomial.

For a large number of particles, we can approximate the solution $x$ to 
\begin{equation}
    x_0 = A+(-1)^A2A\mathrm{e}^{-2A} + \mathcal{O}(A^2\mathrm{e}^{-4A}),
\end{equation}
and assuming that $N\gg1$, then we can approximate to
\begin{equation}
    x_0 = A +\mathcal{O}(A\mathrm{e}^{-2A}).
\end{equation}
Now, using this expression in Eq.~\ref{eq:col_eigvals}, after reindexing the sum, we obtain the general expression
\begin{equation}
    \epsilon_\mathrm{max}(N) = \frac{N!}{A^{2N_p}}\sum_{j=0}^{N_p-2} \frac{(N_p-1-j)}{(N-2N_p+2j)!} A^{2j}    +\mathcal{O}(A\mathrm{e}^{-2A}).
\end{equation}
At this point, it is convenient to define $a_k \equiv A^k/k!$, and the function
\begin{equation}
    T_m(A) = \sum_{k=0}^m (-1)^ka_k = \frac{1}{m!}\int_0^\infty \mathrm{e}^{-t}(t-A)^m dt,
\end{equation}
and taking into account the property
\begin{equation}
    (A-k)a_k = A(a_k-a_{k-1}),
\end{equation}
we can obtain a shorter expression for the eigenvalue. In particular, for $N$ even, doing the transformation $k=2j$, and so $N_p-1-j = (A-k)/2$, then
\begin{align}
        \epsilon_\mathrm{max}(N) &= \frac{1}{2}\frac{N!}{A^{N}}\sum_{k=0}^{A-2} (A-k)\frac{A^k}{k!}   +\mathcal{O}(A\mathrm{e}^{-2A}) \nonumber\\
        &= \frac{1}{2} \frac{N!}{A^{N}}\sum_{k=0}^{A-2} (A-k)a_k   +\mathcal{O}(A\mathrm{e}^{-2A}) \nonumber\\
        &= \frac12\frac{N!}{A^{N-1}}\sum_{k=0}^{A-2}(a_k-a_{k-1}) +\mathcal{O}(A\mathrm{e}^{-2A}) \nonumber\\
        &= \frac12 \frac{N!}{A^{N-1}} T_{A-2}(A) +\mathcal{O}(A\mathrm{e}^{-2A}),
\end{align}
where the sum runs only over the even values of $k$. Similarly, for an odd number of particles, we proceed by replacing $k=2j+1$, and also setting $p-1-j = (A-k)/2$, obtaining the same final result. Still, as the sum runs over odd values of $k$, there is an additional minus sign. As a general result, for even and odd numbers of particles, we can express the maximum eigenvalue as
\begin{equation}
    \epsilon_\mathrm{max}(N) =\frac12 (-1)^A \frac{N!}{A^{N-1}}T_{A-2}(A) +\mathcal{O}(A\mathrm{e}^{-2A}).
\end{equation}
Using the integral representation of $T_m(A)$, we can rewrite $\epsilon_\mathrm{max}$ as 
\begin{align}
    \epsilon_\mathrm{max}(N) =& \frac12 (-1)^{A} \frac{(A+2)!}{A^{A+1}(A-2)!}\int_0^\infty \mathrm{e}^{-t}(t-A)^{A-2} dt \nonumber \\
    =& \frac12 \left(A+2-\frac{1}{A}-\frac{2}{A^2}\right)\int_0^\infty \mathrm{e}^{-t}\left(1-\frac{t}{A}\right)^{A-2}dt.
\end{align}
In the limit $t/A \ll 1$ we can approximate 
\begin{equation}\label{eq:log_aprox}
    (A-2)\log \left(1-\frac{t}{A}\right) = -t +\frac{2t-t^2/2}{A} +\frac{t^2-t^3/3}{A^2} + \mathcal{O}(A^{-3}),
\end{equation}
and by computing 
\begin{align}
    \mathrm{e}^{-t}&\left(1-\frac tA \right)^{A-2} = \mathrm{e}^{-t}\mathrm{e}^{(A-2)\log\left(1-\frac{t}{A}\right)}  \\
    &= \mathrm{e}^{-2t} \mathrm{e}^{\frac{2t-t^2/2}{A} +\frac{t^2-t^3/3}{A^2} + \mathcal{O}(A^{-3})} \nonumber \\
    &= \mathrm{e}^{-2t} \left( 1+\frac{2t-t^2/2}{A} + \frac{3t^2-4t^3/3+t^4/8}{A^2} +\mathcal{O}(A^{-3}) \right). \nonumber
\end{align}
Note that in the regime where the approximation of Eq.~(\ref{eq:log_aprox}) is no longer valid, the exponential makes the integrand vanish. We integrate using $\int_0^\infty t^q\mathrm{e}^{-2t} dt = q!/2^{q+1}$, obtaining 
\begin{equation}
    \epsilon_\mathrm{max}(N) = \frac{A}{4} + \frac{11}{16} + \frac{19}{64A} + \mathcal{O}(A^{-2}),
\end{equation}
and replacing $A=N-2$, we recover the expression Eq.~(\ref{eq:eigval_large_N}).
%
\section{Projection picture for \texorpdfstring{$N=3$}{N=3} and \texorpdfstring{$N=4$}{N=4}}
\label{app:projection_examples}
In this Appendix, we illustrate the projection picture underlying the physical interpretation of the TBDM spectrum and explain why this picture correctly predicts the different degeneracies but does not always yield the explicit eigenvectors directly. 

We use the same convention for the TBDM as in the main text,
\begin{equation}
    \Tr\,\rho^{(2)} = \frac{N(N-1)}{2},
\end{equation}
so that in second quantization we write
\begin{equation}
    \rho^{(2)}
    = \sum_{\vec P}
      \langle \vec P | \Psi_T \rangle
    \langle \Psi_T | \vec P \rangle ,
\end{equation}
where the sum runs over all $(N-2)$-particle occupation states
$|\vec P\rangle$. This expression makes explicit which two-body
states remain after tracing out the other $N-2$ particles.

We define $\lambda^{(N)}_k$ as the $k$-th eigenvalue of the one-body density matrix of the $ N$- particle system.
\subsection{\texorpdfstring{$N=3$}{N=3}}

For $N=3$, one traces out a single particle. Using the odd-$N$ FTG state,
\begin{equation}
    |\Psi_T\rangle
    = \sum_k \sqrt{\lambda_k^{(3)}}
    \,|1_0,\ldots,1_{k+},1_{k-},\ldots\rangle ,
\end{equation}
the trace can be decomposed as
\begin{align}
    \rho^{(2)}
    ={}& \langle 1_0 | \Psi_T \rangle
        \langle \Psi_T | 1_0 \rangle
    + \sum_k \langle 1_{k+} | \Psi_T \rangle
        \langle \Psi_T | 1_{k+} \rangle
    \nonumber\\
    &+ \sum_k \langle 1_{k-} | \Psi_T \rangle
        \langle \Psi_T | 1_{k-} \rangle .
\end{align}
Evaluating each term gives
\begin{equation}
    \langle 1_0 | \Psi_T \rangle
    = \sum_k \sqrt{\lambda_k^{(3)}}\, |1_{k+}\,1_{k-}\rangle ,
\end{equation}
\begin{align}
    \langle 1_{k+} | \Psi_T \rangle
    &= \sqrt{\lambda_k^{(3)}}\, |1_0\,1_{k-}\rangle , \\
    \langle 1_{k-} | \Psi_T \rangle
    &= \sqrt{\lambda_k^{(3)}}\, |1_0\,1_{k+}\rangle .
\end{align}
Therefore,
\begin{align}
    \rho^{(2)}
    ={}&  |\alpha\rangle\langle\alpha|
    + \sum_k \lambda_k^{(3)}
    |1_0\,1_{k+}\rangle\langle 1_0\,1_{k+}|
    \nonumber\\
    &+ \sum_k \lambda_k^{(3)}
    |1_0\,1_{k-}\rangle\langle 1_0\,1_{k-}| ,
\end{align}
where
\begin{equation}
    |\alpha\rangle
    = \sum_k \sqrt{\lambda_k^{(3)}}\, |1_{k+}\,1_{k-}\rangle .
\end{equation}

In this case, all the states appearing above are mutually orthogonal.
Hence, this is already the spectral decomposition of the TBDM. The
collective state $|\alpha\rangle$ gives the singlet, while the states
$|1_0\,1_{k\pm}\rangle$ give the doublets.

\subsection{\texorpdfstring{$N=4$}{N=4}}

For $N=4$, one traces out two particles. Using the even-$N$ FTG state,
\begin{equation}
    |\Psi_T\rangle
    = \sum_{k_1<k_2} \sqrt{p_4(k_1,k_2)}\,
    |1_{k_1+},1_{k_1-},1_{k_2+},1_{k_2-}\rangle ,
\end{equation}
the trace can be written by separating the different possibilities
for the two traced particles. First, the terms in which the two traced
particles belong to the same pair:
\begin{equation}
    \sum_{k_3}
    \langle 1_{k_3+}\,1_{k_3-} | \Psi_T \rangle
    \langle \Psi_T | 1_{k_3+}\,1_{k_3-} \rangle .
\end{equation}
Second, the terms in which the traced particles belong to two
different pairs:
\begin{multline}
    \sum_{k_3<k_4}
    \langle 1_{k_3+}\,1_{k_4+} | \Psi_T \rangle
    \langle \Psi_T | 1_{k_3+}\,1_{k_4+} \rangle \\
    + \sum_{k_3<k_4}
    \langle 1_{k_3+}\,1_{k_4-} | \Psi_T \rangle
    \langle \Psi_T | 1_{k_3+}\,1_{k_4-} \rangle \\
    + \sum_{k_3<k_4}
    \langle 1_{k_3-}\,1_{k_4+} | \Psi_T \rangle
    \langle \Psi_T | 1_{k_3-}\,1_{k_4+} \rangle \\
    + \sum_{k_3<k_4}
    \langle 1_{k_3-}\,1_{k_4-} | \Psi_T \rangle
    \langle \Psi_T | 1_{k_3-}\,1_{k_4-} \rangle .
\end{multline}
Thus,
\begin{align}
    \rho^{(2)}
    = \Bigg[
    &\sum_{k_3}
    \langle 1_{k_3+}\,1_{k_3-} | \Psi_T \rangle
    \langle \Psi_T | 1_{k_3+}\,1_{k_3-} \rangle
    \nonumber\\
    &+
    \sum_{k_3<k_4}
    \sum_{\sigma,\sigma'=\pm}
    \langle 1_{k_3\sigma}\,1_{k_4\sigma'} | \Psi_T \rangle
    \langle \Psi_T | 1_{k_3\sigma}\,1_{k_4\sigma'} \rangle
    \Bigg].
\end{align}

Evaluating the first type of term gives
\begin{align}
    \langle 1_{k_3+}\,1_{k_3-} | \Psi_T \rangle
    ={}& \sum_{k\neq k_3} \sqrt{p_4(k,k_3)}\, |1_{k+}\,1_{k-}\rangle
    \nonumber\\
    \equiv{}& \sqrt{\lambda_{k_3}^{(4)}}\, |\alpha_{k_3}\rangle ,
\end{align}
with
\begin{equation}
    |\alpha_{k_3}\rangle
    = \frac{1}{\sqrt{\lambda_{k_3}^{(4)}}}
    \sum_{k\neq k_3} \sqrt{p_4(k,k_3)}\, |1_{k+}\,1_{k-}\rangle ,
\end{equation}
and
\begin{equation}
    \lambda_{k_3}^{(4)}
    = \sum_{k\neq k_3} p_4(k,k_3).
\end{equation}

For the second type of term, if the two traced particles belong to
different pairs, the remaining two-body state is fixed. For instance,
\begin{equation}
    \langle 1_{k_3+}\,1_{k_4+} | \Psi_T \rangle
    = \sqrt{p_4(k_3,k_4)}\, |1_{k_3-}\,1_{k_4-}\rangle ,
\end{equation}
and analogously for the other three sign combinations. Hence, the TBDM becomes
\begin{align}
    \rho^{(2)}
    ={}&  \sum_{k_3} \lambda_{k_3}^{(4)}
    |\alpha_{k_3}\rangle\langle \alpha_{k_3}|
    \nonumber\\
    &+ \sum_{k_3<k_4} p_4(k_3,k_4)
    \sum_{\sigma,\sigma'=\pm}
    |1_{k_3\sigma}\,1_{k_4\sigma'}\rangle
    \langle 1_{k_3\sigma}\,1_{k_4\sigma'}| .
\end{align}

At this point, the physical interpretation is clear. The states 
$|1_{k_3\sigma}\,1_{k_4\sigma'}\rangle$
are the quadruplets: they come from tracing out two particles
belonging to different pairs, and are orthonormal. Therefore the
projection picture correctly predicts both their existence and their
exact form.

The singlet sector is subtler. The states $|\alpha_{k_3}\rangle$ are
collective states, since they are superpositions of all pair modes
$(k,-k)$ compatible with the traced pair $k_3$. However, they are not
orthogonal in general:
\begin{align}
    \langle \alpha_{k_3} | \alpha_{k_4} \rangle
    ={}& \frac{1}{\sqrt{\lambda_{k_3}^{(4)}\lambda_{k_4}^{(4)}}}
    \sum_{k\neq k_3,k_4}
    \sqrt{p_4(k,k_3)p_4(k,k_4)}
    \nonumber\\
    &\neq 0 .
\end{align}
Therefore, unlike the $N=3$ case, the expression above is not yet the
spectral decomposition of the TBDM in the singlet sector. One must
still diagonalize this non-orthogonal block to obtain the exact singlet
eigenvectors and eigenvalues.
\bibliography{biblio_FTG}
\end{document}